\documentclass[twocolumn,superscriptaddress]{revtex4}
\usepackage{color}
\usepackage{graphics,graphicx,epsfig}
\usepackage{amssymb,amsfonts,amsmath}
\include{epsf}
\usepackage{ifthen}

\newcommand{\beqn}{\begin{eqnarray}}
\newcommand{\eeqn}{\end{eqnarray}}
\newcommand{\beq}{\begin{equation}}
\newcommand{\eeq}{\end{equation}}

\usepackage{ifthen}

\begin{document}
\title{Inferring processes underlying B-cell repertoire diversity. }


\author{Yuval Elhanati${}^{*}$}
\affiliation{Laboratoire de physique th\'eorique, UMR8549, CNRS and \'Ecole normale sup\'erieure, 24, rue Lhomond, 75005 Paris, France}
\thanks{These two authors contributed equally}
\author{Zachary Sethna${}^{*}$}
\affiliation{Joseph Henry Laboratories, Princeton University, Princeton, New Jersey 08544 USA} \author{Quentin Marcou}
\affiliation{Laboratoire de physique th\'eorique, UMR8549, CNRS and \'Ecole normale sup\'erieure, 24, rue Lhomond, 75005 Paris, France}
\author{Curtis G. Callan Jr.}
\affiliation{Joseph Henry Laboratories, Princeton University, Princeton, New Jersey 08544 USA}
\author{Thierry Mora}
\affiliation{Laboratoire de physique statistique, UMR8550, CNRS and \'Ecole normale sup\'erieure, 24, rue Lhomond, 75005 Paris, France}
\author{Aleksandra M. Walczak}
\affiliation{Laboratoire de physique th\'eorique, UMR8549, CNRS and \'Ecole normale sup\'erieure, 24, rue Lhomond, 75005 Paris, France}

\date{\today}

\begin{abstract}

We quantify the VDJ recombination and somatic hypermutation processes in human B-cells using probabilistic inference methods on high-throughput DNA sequence repertoires of human B-cell receptor heavy chains. Our analysis captures the statistical properties of the naive repertoire, first after its initial generation via VDJ recombination and then after selection for functionality. We also infer statistical properties of the somatic hypermutation machinery (exclusive of subsequent effects of selection). Our main results are the following: the B-cell repertoire is substantially more diverse than T-cell repertoires, due to longer junctional insertions; sequences that pass initial selection are distinguished by having a higher probability of being generated in a VDJ recombination event; somatic hypermutations have a non-uniform distribution along the V gene that is well explained by an independent site model for the sequence context around the hypermutation site.

\end{abstract}

\maketitle

\section{Background}

Along with T-cells, B-cells contribute to the large diversity of immune cells that specifically recognize antigens. The diversity of the B-cell repertoire is encoded in the different amino acid composition of B cell receptors (BCRs) expressed on the surface of these cells. These receptors are formed during the VDJ recombination process in the bone marrow. Before these cells leave for the periphery, they are initially selected for functionality. Later, they undergo further selection depending on their ability to recognize foreign antigen. Additionally, unlike T-cells, B-cell receptor sequences are subject to point hypermutations during the proliferation that follows successful recognition of an antigen  \cite{Teng07}. These hypermutations are selected for antigen binding through the process of affinity maturation. Apart from the possibility of hypermutations,  the generation and selection processes are very similar in B- and T-cells, and involve common enzymes and pathways \cite{Janeway}. Recent advances in sequencing technologies \cite{Six2013,Robins2013646} make it possible to obtain copious data on immune cell receptor sequences. We work with large samples of human B-cell receptors heavy chain \footnote{Data from Harlan Robins group, available at {\tt http://physics.princeton.edu/\textasciitilde ccallan/BCRPaper}.} 
and apply advanced statistical techniques (developed in \cite{Murugan02102012, Elhanati08072014} and previously used to describe T-cell repertoires) to accurately quantify the processes that shape B-cell repertoire diversity -- generation, selection and hypermutations.

Many characteristics of B-cell repertoires have previously been described using a variety of methods. 
Gene usage was studied by both immunoscope techniques \cite{Lim01012008, Foreman08} and single cell PCR \cite{Brezinschek97}, and the variable length distributions of the CDR3 region were reported \cite{Rock01011994, Wu93}. 
A number of studies have characterized the effects of B-cell selection, reporting length reduction and selection against hydrophobicity as dominant features \cite{Shiokawa99, Ivanov15062005, Wardemann05092003, Uduman01072011,Harlan12,Raaphorst01101997,Meffre2001}. Considerable attention has also been given to quantifying hypermutations and disentangling them from site specific selection, either by comparing synonymous and non-synonymous mutations, or functional and non-functional rearrangements \cite{Matsen14, Yaari13, Shapiro01071999, Betz15031993, Cowell15022000, Kepler14, Lossos01112000, Dunn-Walters01031998}. Other studies used lineage trees to describe this mutation-selection process \cite{Kepler14, SteimanShimony2006130}.
Recently, high-throughput sequencing data combined with sophisticated inference techniques have been used to study selection in the affinity maturation process \cite{Matsen14} and the statistics of synonymous hypermutation profiles  \cite{Yaari13}.

Yet a comprehensive quantitative description of the entire generation, selection and hypermutation process of the heavy chain repertoire is still lacking. Here we self-consistently model all parts of these interlinked processes.
VDJ recombination is based on a combinatoric process in which V,  D and J genes are chosen from a number of genomic templates and joined together, with additional base pair insertions and deletions at the VD and DJ junctions leading to further randomness in the final sequence \cite{Janeway}. In the case of antigen-experienced cells, receptor sequences further undergo random somatic hypermutations. A difficulty that arises in analyzing receptor sequences is that a given sequence can be produced in many ways. This complication makes it impossible to unambiguously retrace the steps (V, D, J gene choices, deletions and insertions at the junctions, hypermutations) that have led to its generation. Our method circumvents this difficulty by employing a probabilistic framework that exploits large sequence data sets to accurately infer the statistical properties of the three processes that are central to the generation and evolution of B cell receptors---VDJ recombination, functional selection and hypermutations. This analysis allows us to quantify the theoretical diversity of these sequences.
Our results suggest that, as in the case of T-cell receptors, the VDJ recombination process is biased towards sequences that are likely to pass functional selection. We also show that the diversity of the human B-cell repertoire is significantly reduced during the initial functional selection step. 
Our probabilistic framework also allows us to account for the statistics of hypermutations, which are well described by an additive DNA context dependent binding model.



\section{Analysis approach}

We analysed Illumina reads of B-cell receptor (BCR) DNA sequences from human blood samples taken from two individuals (labeled A and B) \cite{Harlan12}. Cells from each sample were 
sorted 
into naive and memory sub-samples. The variable region of the B-cell receptor gene was amplified using sequence specific PCR primers resulting in 130 bp DNA sequence reads anchored on a conserved sequence within the J gene (data from H. Robins, whom we thank for sharing it with us).

\begin{figure}
\begin{center}
\includegraphics[width=\linewidth]{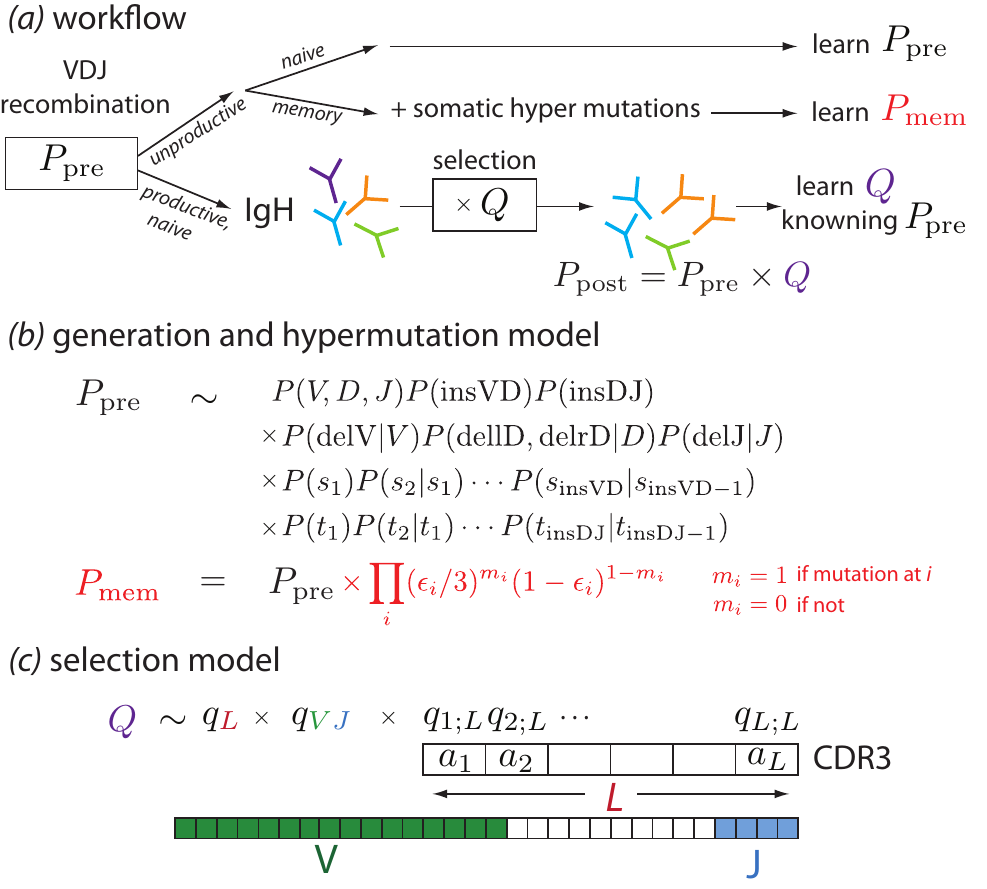}
\caption{({\bf a}) B cell receptor heavy chain sequences are formed during VDJ recombination according to a probability distribution $P_{pre}$ that we infer from the unproductive naive sequence repertoire. The unproductive memory repertoire is used to infer the rate and sequence dependence of somatic hypermutation. Productive sequences are selected for entry into the naive peripheral repertoire with a sequence-dependent factor $Q$, resulting in the observed distribution of receptor sequences $P_{post}$ . ({\bf b})  Recombined sequences arise via a scenario  involving independent choices of which gene segments to recombine as well as of numbers of deletions and insertions. The probability distribution of these choices is not known unambiguously from the observed sequences and is estimated probabilistically in an iterative procedure. ({\bf c}) The selection factor $Q$ is assumed to be a product of factors for V and J gene choice together with factors $q_{i,L}(a)$ for the choice of the specific amino acid $a$ at each position $i$ in a CDR3 region of length $L$. These factors are determined from the naive productive sequence repertoire by an iterative procedure.}
\label{analysis_cartoon}
\end{center}
\end{figure}

The VDJ recombination process is not guaranteed to produce in-frame sequences or, even when sequences are in frame, functional proteins. If the receptor gene from the initially rearranged chromosome is not functional, the second chromosome may be rearranged. If this second recombination event leads to a functional receptor, the cell has two rearranged chromosomes---one functional and expressed, and the other one silenced by allelic exclusion. As a result, the DNA sequence dataset we analyzed contains a large fraction of {\em non-productive} sequences, which are either out-of-frame or contain a stop codon. These sequences experienced no selection and owe their survival to the receptor expressed by the other chromosome. 
For this reason they provide us with the raw, unselected product of the generation process.
We used such out-of-frame sequences from the naive subsample to infer the statistics of the VDJ recombination process, and the out-of-frame sequences from the memory subsample to learn the statistics of hypermutations. McCoy et al \cite{Matsen14} previously exploited these differences between in- and out-of-frame sequences in human BCR memory repertoire analysis. The naive {\em productive} sequences (in frame and with no stop codon) are expected to have passed a selection process before being admitted to the periphery (henceforth called {\em initial} selection, to distinguish it from selection following a recognition event). 
 We used this subsample to learn these selective forces acting on the amino-acid, by comparing how their statistics differ from the raw product of VDJ recombination learned from the naive out-of-frame sequences. Fig.\,\ref{analysis_cartoon}a summarizes the analysis workflow, and emphasizes how the three main processes underlying sequence diversity---VDJ recombination, initial selection, hypermutations---are infered using three subsamples of the sequences.
A typical subsample used in our analysis had 
$\sim200,000$ unique sequences.



As we stressed above, one sequence can be the result of a number of scenarios that include different initial gene choices, followed by variable numbers of deleted and inserted base pairs. This problem requires a probabilistic description of the generation process, which sums over all the different possible scenarios given their relative weighted contributions for producing a given sequence. Each scenario's probability $P_{\rm pre}$ is calculated using a generation model of the form shown in Fig.\,\ref{analysis_cartoon}b. 
In brief, the various factors account for the probabilities of uncorrelated events leading to a specific VDJ rearrangement: choice of which gene segments to recombine $P(V,D,J)$, choice of number of nucleotides to insert in a VD or DJ joint $P({\rm insVD})$ and $P({\rm insDJ})$, probability of number of deletions from all four ends of the V, D and J genes at the junctions $P({\rm delV}|V)$, $P({\rm dellD},{\rm delrD}|D)$ and $P({\rm delJ}|J)$,
as well as factors to account for unequal nucleotide preference in the inserted sequences. 
Since the recombination machinery is the same for B-cells and T-cells, and this model structure captured all the correlations present in T-cell data \cite{Murugan02102012}, we expect that it should also capture all correlations present in the B-cell sequence data; we have verified that this is the case.
In \cite{Murugan02102012} we described an Expectation-Maximisation method to self-consistently solve for the component probability distributions in $P_{\rm pre}$ by maximising the likelihood of the set of observed sequences and we apply the same method here.
This method is applied to the naive unproductive sequence repertoires, which result from the raw generation process.

B-cell receptor sequences also acquire point hypermutations during proliferation, which occur with probability $\epsilon_i$ at position $i$ on the sequence. We use the non-productive memory sequences to learn $P_{\rm mem}$, which is the product of $P_{\rm pre}$ and the probability of a given combination of mutations (given in red in Fig.~\ref{analysis_cartoon}b). As in the case of non-productive naive sequences, we reasoned that these sequences were not selected at any point (either before or after hypermutations), and that their statistics should reflect the raw product of VDJ recombination followed by random mutations.




\begin{figure}
\begin{center}
\includegraphics[width=\linewidth]{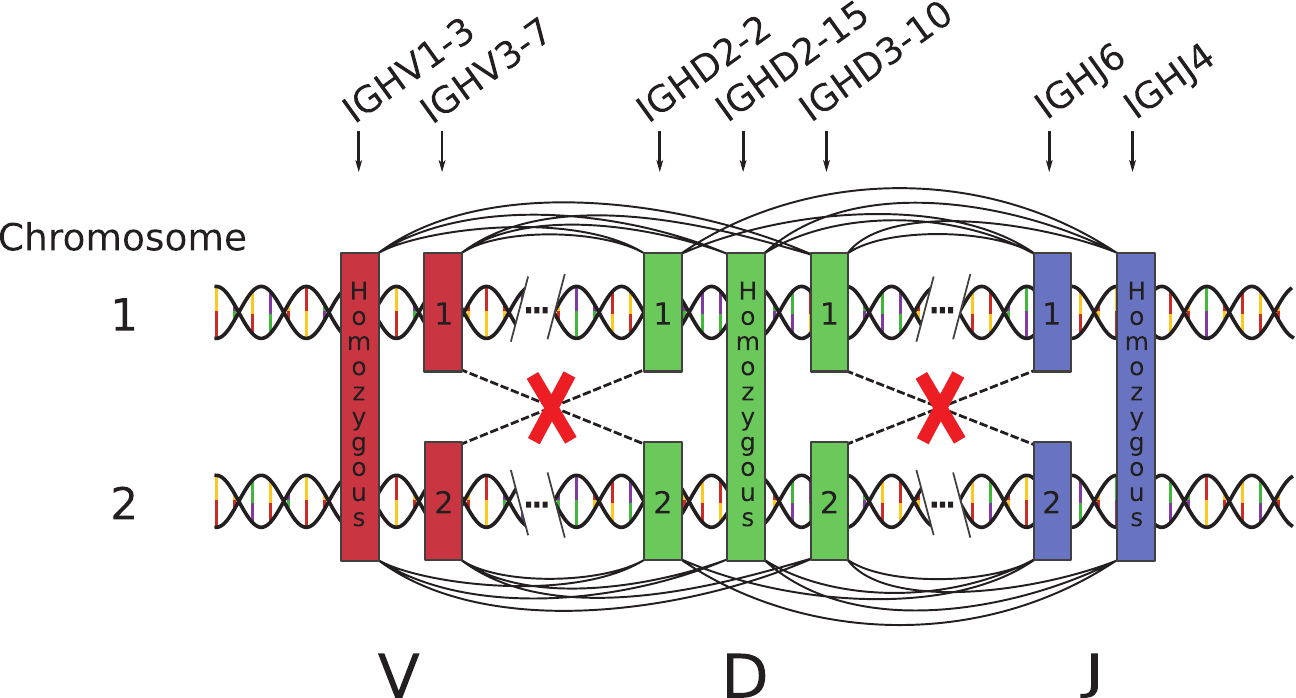}
\caption{The organization of heterozygous genes into chromosomes can be probabilistically determined. Every recombination event ties together a V, a D, and a J gene, as indicated by the arcs drawn above and below the two chromosomes. Links that recombine alleles on different chromosomes are forbidden (red crosses). Our method gives the probability $P(V,D,J)$ of all possible linkages between three genes (distinguishing between alleles of the same gene), but does not address how the various alleles are grouped on chromosomes. We find the best chromosomal segregation by minimizing the sum of all terms in $P(V,D,J)$ that contain forbidden links (red crosses).}
\label{allele_seg}
\end{center}
\end{figure}

Using the (hypermutation-free) generation model as a starting point, we infer selection factors $Q$ acting on each sequence in the naive repertoire, where $Q$ is defined as the fold increase of the probability to see a particular sequence in the functional repertoire (naive, productive) compared to the previously learned generation probability: $Q= P_{\rm post}/P_{\rm pre}$. To infer those factors, we use a factorized model (Fig.\,\ref{analysis_cartoon}c), where we assume that selection acts independently on the V and J gene choice (through factor $q_{VJ}$), the length $L$ of the CDR3 sequence (through factor $q_L$), and on each of the amino acids $a_i$ at positions $1\leq i\leq L$ between the conserved Cysteine near the end of the V gene and the conserved Tryptophan within the J gene  (through factors $q_{i;L}(a_i)$). We use an expectation maximization procedure to update the selection factors until convergence, as
previously described in \cite{Elhanati08072014}.

Our inference of recombination scenarios for individual reads requires accurate knowledge of the germline sequence of all the V, D, and J genes. These genes have several alleles, and an accurate accounting of sequencing errors and somatic hypermutation events requires knowing which alleles are present on the two chromosomes of each individual.
The existing databases do not provide such information, but list all alleles that have been detected, together with an estimate of the population frequency of these alleles \cite{IMGT}. 
To address this problem, we have developed a method for identifying alleles directly from the sequence data for each individual. 
By accumulating patterns of mismatches between reads and database gene sequences that occur too often to be attributed to chance errors, we can detect the presence of alleles for each gene. This procedure usually identifies at most two significant alleles needed to account for a given individual's reads (the majority of genes are homozygous, but a significant number are not). In summary, we infer specific alleles for each individual, and then use that individual's specific alleles when inferring generative models or selection models. This refinement of the genetic information yields a much improved accounting of sequencing errors and hypermutation events.

Given a generative model inferred using this allelic data, we are able to make a probabilistic assignment of alleles to two chromosomes for each individual. The model includes a joint usage probability distribution $P(V,D,J)$ of the three sets of genes/alleles (in effect treating alleles as separate genes). Since VDJ rearrangement happens on a single chromosome, the probability of recombining a heterozygous V allele with a heterozygous D allele on different chromosomes should be zero, up to assignment errors (Fig.\,\ref{allele_seg}). 
To reconstruct the two chromosomes, we consider all possible associations of a chromosome to each heterozygous allele. We then compute the likelihood of each chromosomal association as the sum of the joint probability $P(V,D,J)$  over all V,D,J combinations that are associated with the same chromosome, and simply choose the chromosomal association that maximizes this likelihood. In this way we found a chromosomal organization for the two individuals that accounted for about 90\% of all sequences.
We can also evaluate the usage probability of the two chromosomes identified using this procedure. For both individuals, it was consistent with equal usage probability between the two chromosomes, within errors.

\section{Results}

\subsection{Distribution of recombination events}

\begin{figure}
\begin{center}
\includegraphics[width=\linewidth]{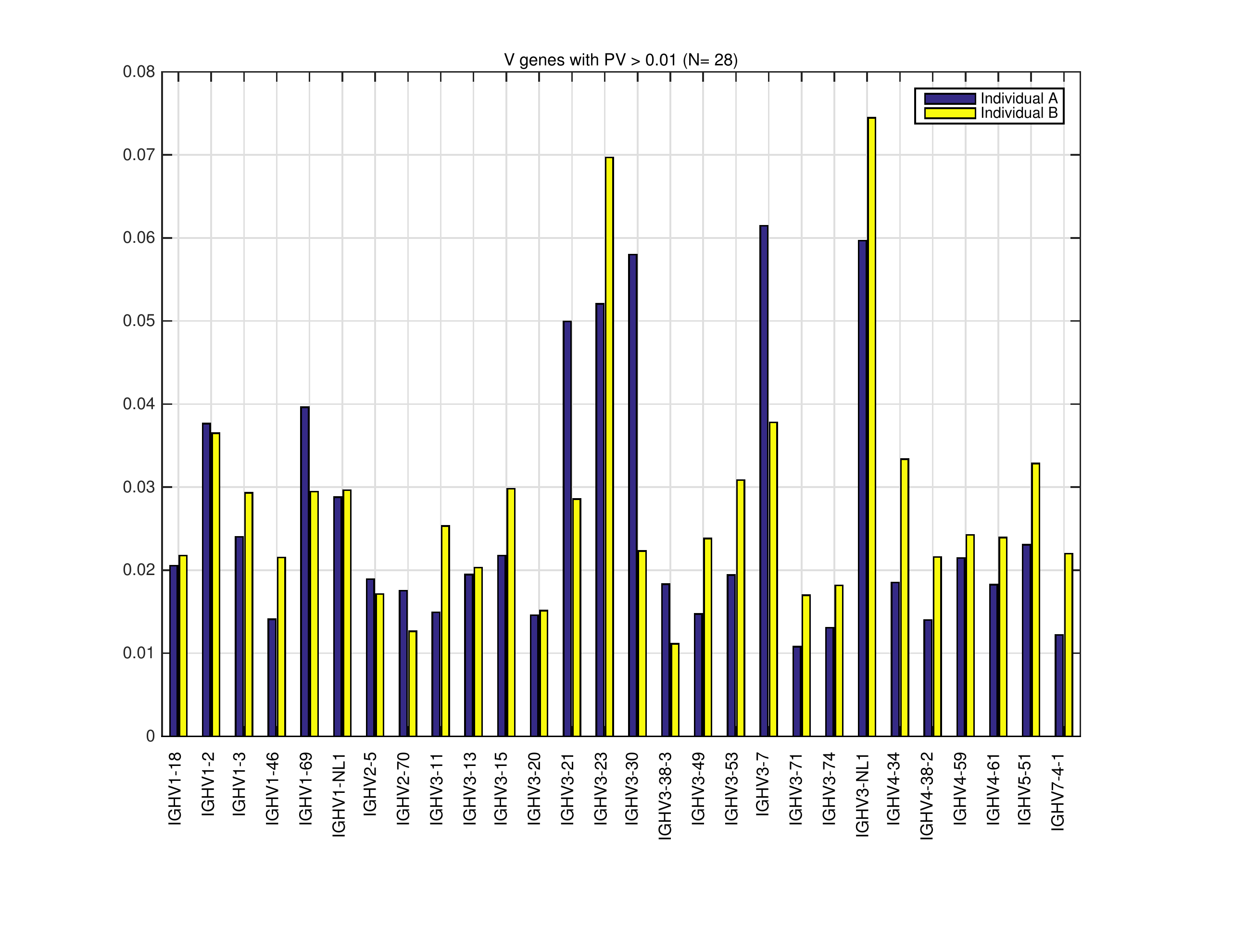}
\caption{V gene usage frequencies in VDJ recombination as inferred from naive nonproductive sequence repertoires for individuals A and B. Usage probability varies widely between genes and this plot displays only those V genes with usage frequency greater than $.01$ for both individuals. Less than half of all V genes pass this filter. The variation between individuals, even for genes with the highest usage rates, is quite substantial.}
\label{S1}
\end{center}
\end{figure}
\begin{figure}
\begin{center}
\includegraphics[width=\linewidth]{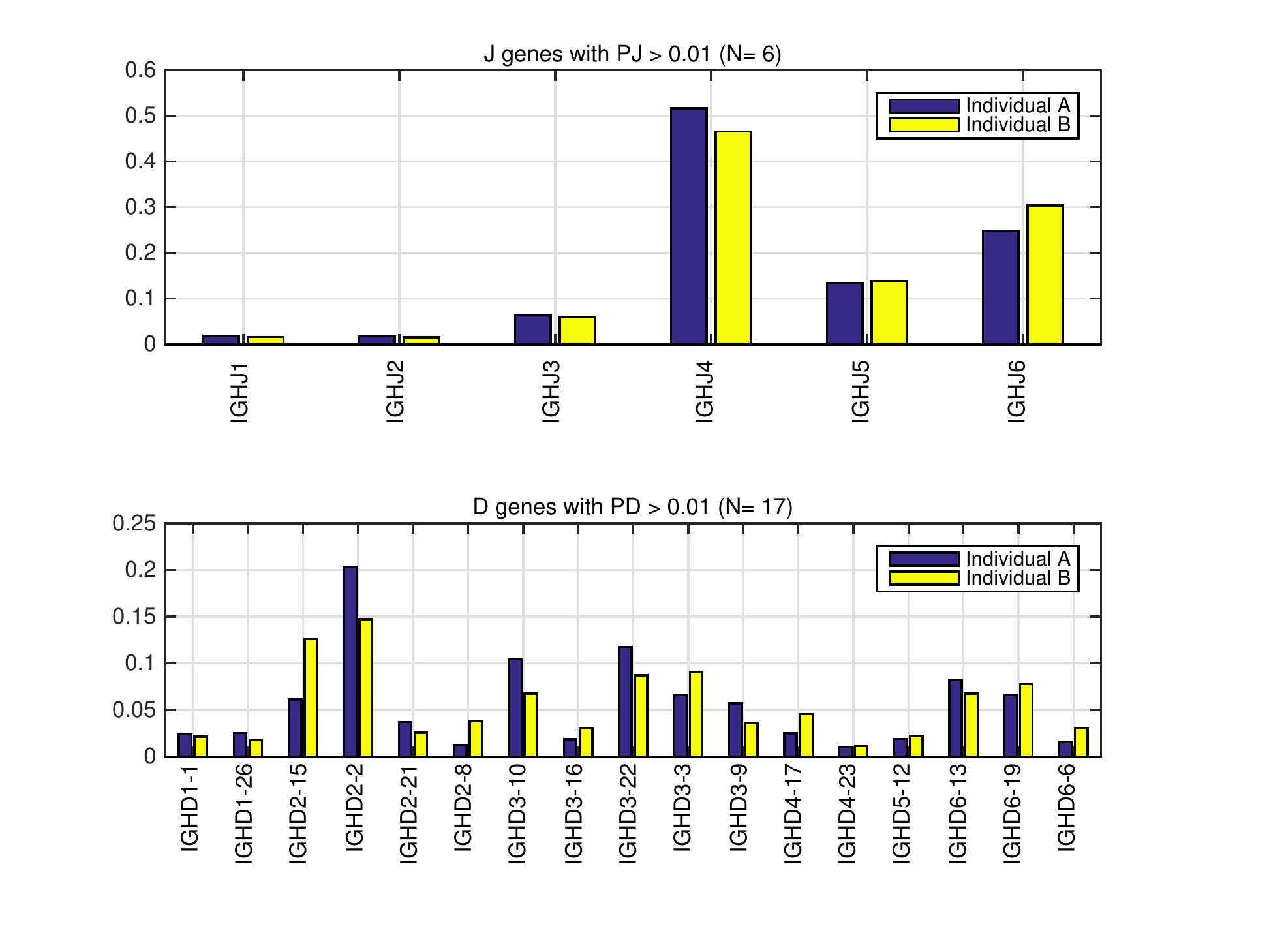}
\caption{D and J gene usage frequencies in VDJ recombination as inferred from naive nonproductive sequence repertoires for individuals A and B. Once again the plot only displays genes with usage frequency greater than $.01$ for both individuals. Less that half of the D genes (but all of the J genes) pass this filter. The variation between individuals of D gene usage (but not of J gene usage) is quite substantial.}
\label{S2}
\end{center}
\end{figure}

The raw distribution of recombination events $P_{\rm pre}$ before any selection takes place was inferred from the naive, non-productive sequence dataset,
as explained above (Fig.\,\ref{analysis_cartoon}). This inference procedure yielded the VDJ gene usage distribution, as well as the distributions of insertions and deletions, and the frequency of inserted nucleotides. Similarly to T-cell generation, the VJ gene usage (Figs.\,\ref{S1} and \ref{S2}) is strongly non-uniform.

\begin{figure}
\begin{center}
\includegraphics[width=\linewidth]{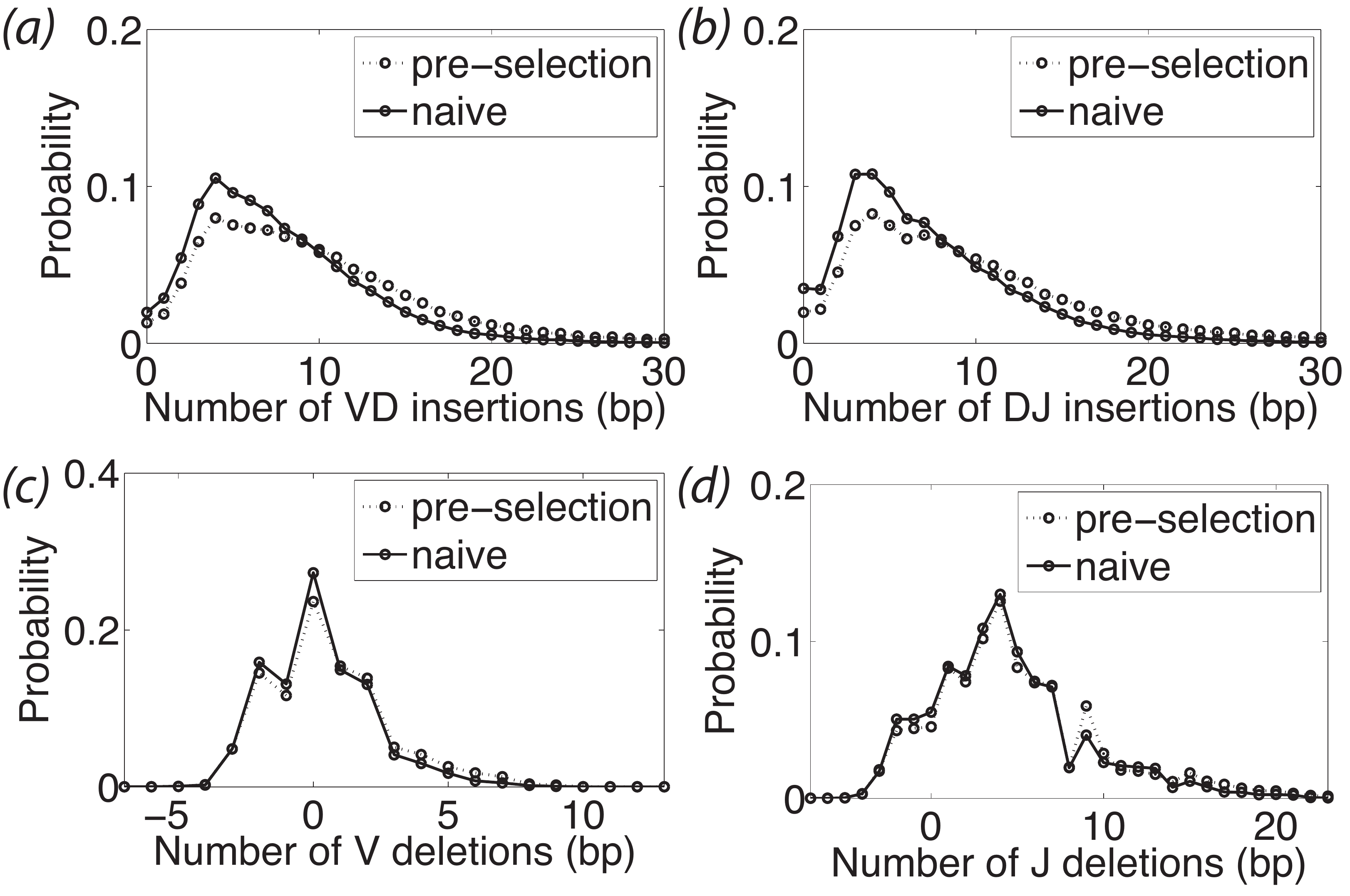}
\caption{Distributions of insertions and deletions for the pre- and post-selection repertoires. The top two panels ({\bf a},{\bf b}) show the distribution of numbers of nucleotide insertions in the DJ and VD joints. These distributions are independent of the identities of the genes on either side of the junction, and the VD and DJ insertions are very similar. The selection process that acts on going from the primitively generated to the naive repertoire causes the mean number of insertions to decrease significantly. The bottom two panels ({\bf c},{\bf d}) show the distribution of deletions from the V and J genes (negative deletions account for palindromic insertions). Deletions are gene-dependent and the plots show the deletion profile averaged over all genes (gene-dependent profiles are shown in Fig.\,S4). Selection has little effect on deletion profiles.}
\label{RecomEvDist}
\end{center}
\end{figure}

Fig.\,\ref{RecomEvDist} shows the distributions of the number of inserted nucleotides between the V and D genes (Fig.\,\ref{RecomEvDist}a), or D and J genes (Fig.\,\ref{RecomEvDist}b), averaged over both individuals. The figure shows both the distributions after generation, {\em i.e.} before any selection, as inferred from the non-productive sequences and, 
for comparison, the same distributions for the productive sequences in the naive repertoire, {\em i.e.} after the initial selection process.
They have similar forms---wide distributions with exponential tails. The effect of selection is to favour sequences with less insertions, thus reducing CDR3 length.

\begin{figure}
\begin{center}
\includegraphics[width=\linewidth]{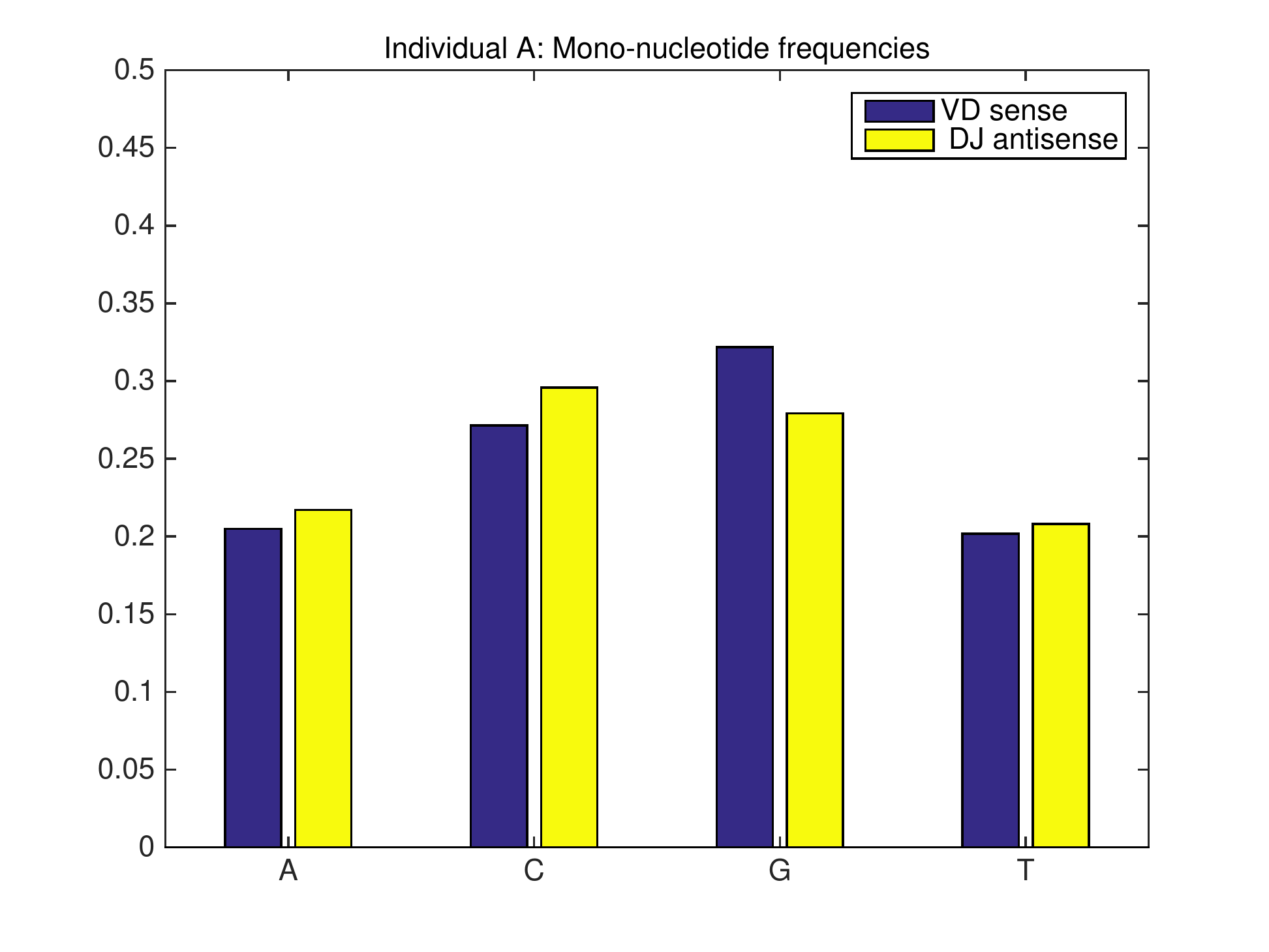}
\includegraphics[width=\linewidth]{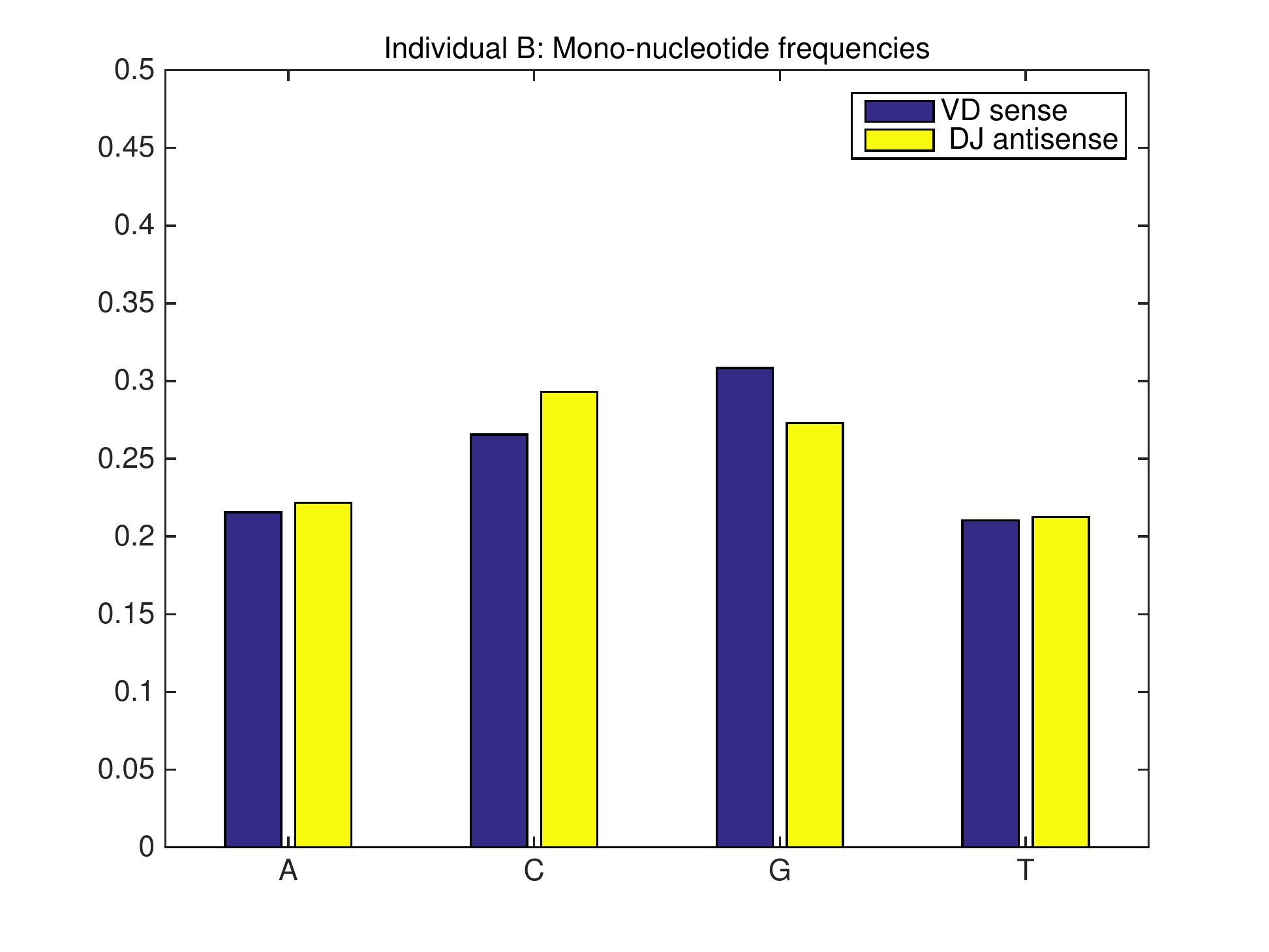}
\caption{Mononucleotide utilization frequency for insertions. Top panel -  individual A, Bottom panel -  individual B. A simple mononucleotide Markov model for generating VD and VJ insertions does not describe their sequence statistics with complete accuracy.}
\label{S5}
\end{center}
\end{figure}

\begin{figure}
\begin{center}
\includegraphics[width=\linewidth]{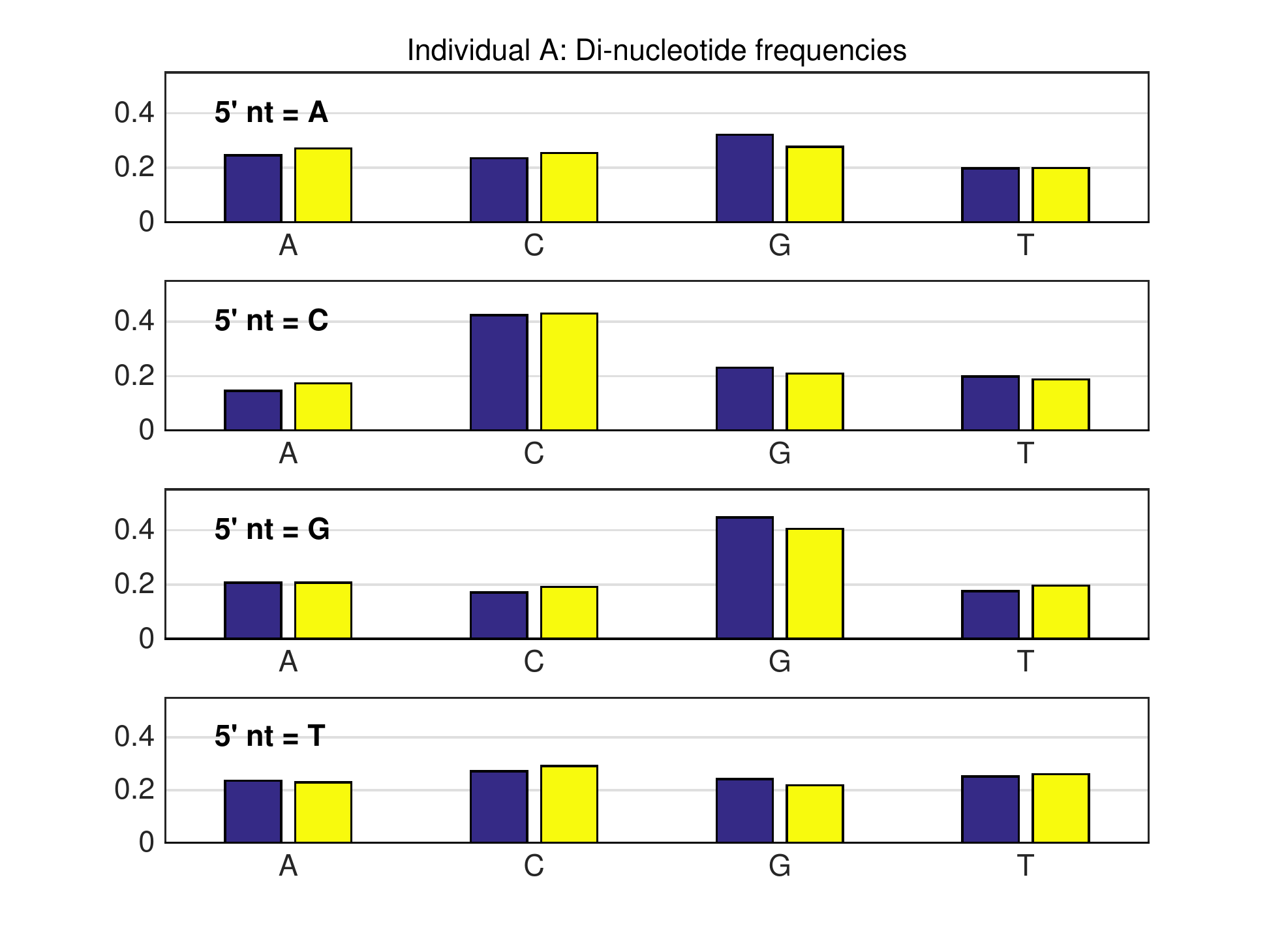}
\includegraphics[width=\linewidth]{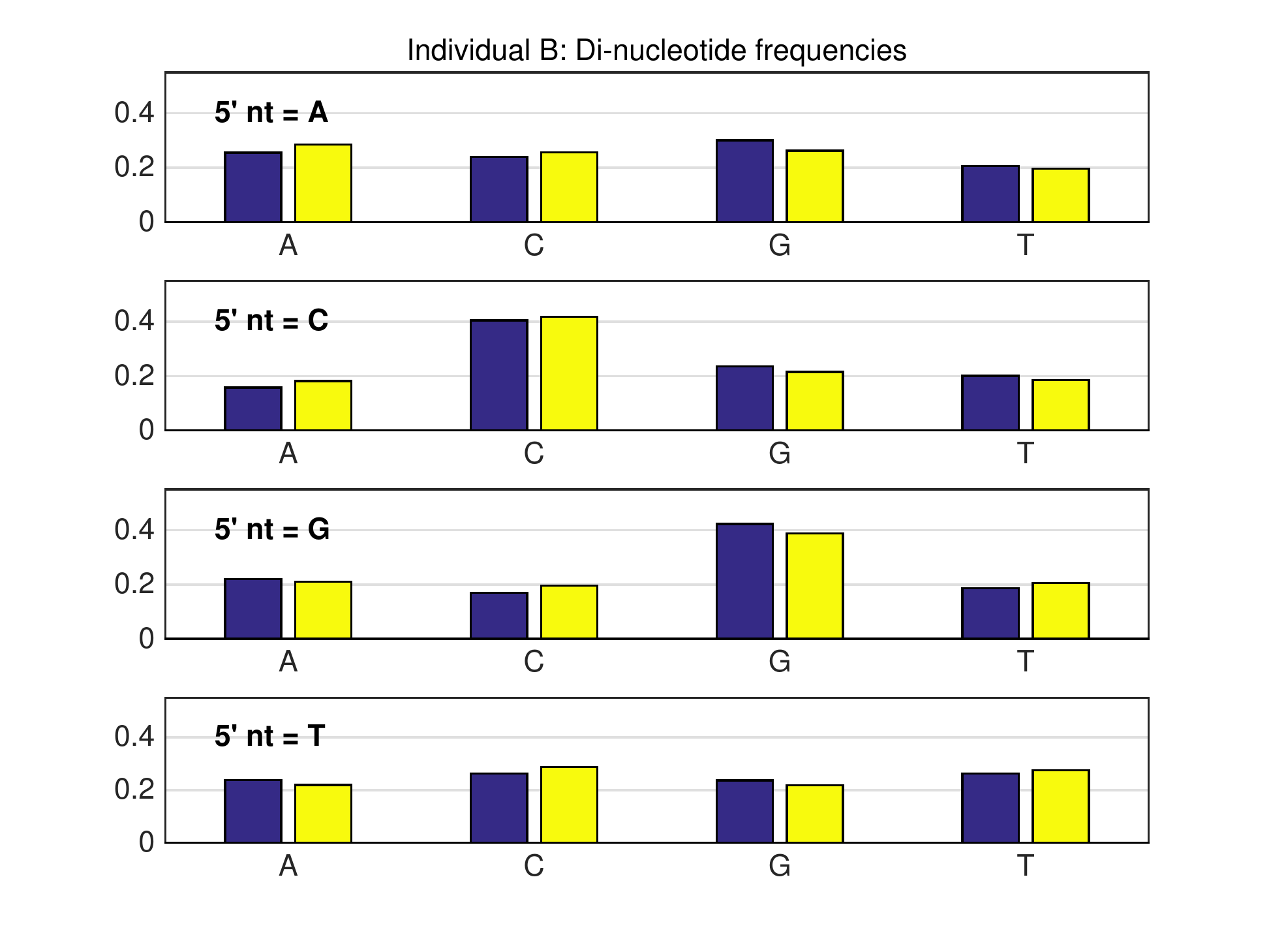}
\caption{Dinucleotide frequencies of VD andDJ insertions. Top panel -  individual A, Bottom panel -  individual B.  A common dinucleotide Markov model accurately describes the sequence statistics of both VD and DJ insertions (reading VD sequences from the DNA top strand and the DJ insertions from the bottom strand). }
\label{S6}
\end{center}
\end{figure}

\begin{figure*}
\begin{center}
\includegraphics[width=\linewidth]{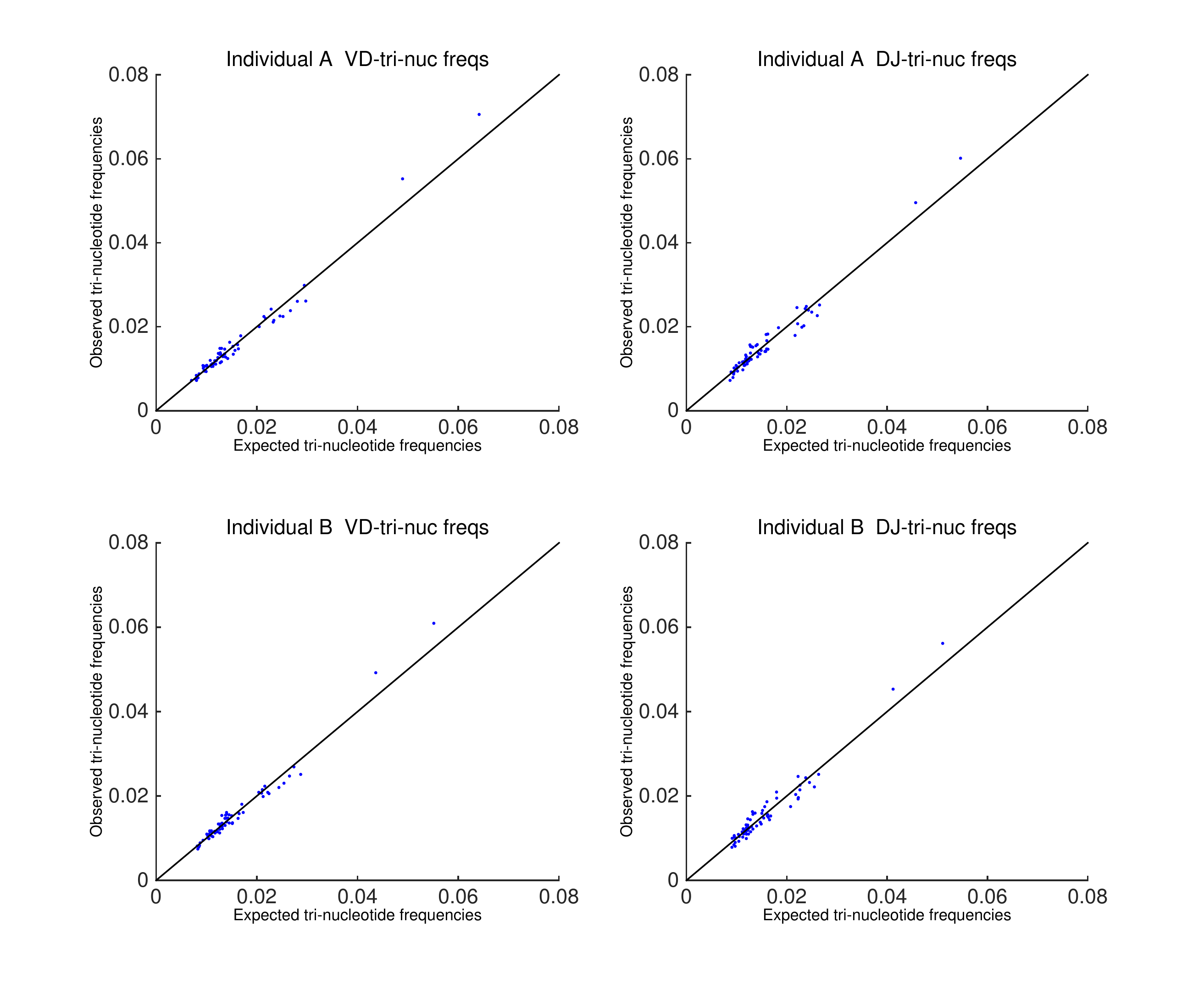}
\caption{Scatter plot of observed trinucleotide frequencies in VD and DJ insertions against their frequencies as predicted by the dinucleotide Markov model. The plot clearly indicates that three-base correlations are accurately reproduced by the dinucleotide Markov model for both individuals and both types of insertion.}
\label{S7}
\end{center}
\end{figure*}

The identity of the nucleotides inserted during the generation process are well described by a dinucleotide Markov model (Fig.\,\ref{S5}-\ref{S7}). 
As was observed for T-cell receptors \cite{Murugan02102012}, the profile for the VD insertions on the sense strand correlates very well with the JD insertions on the anti-sense strand.

\begin{figure*}
\begin{center}
\includegraphics[width=\linewidth]{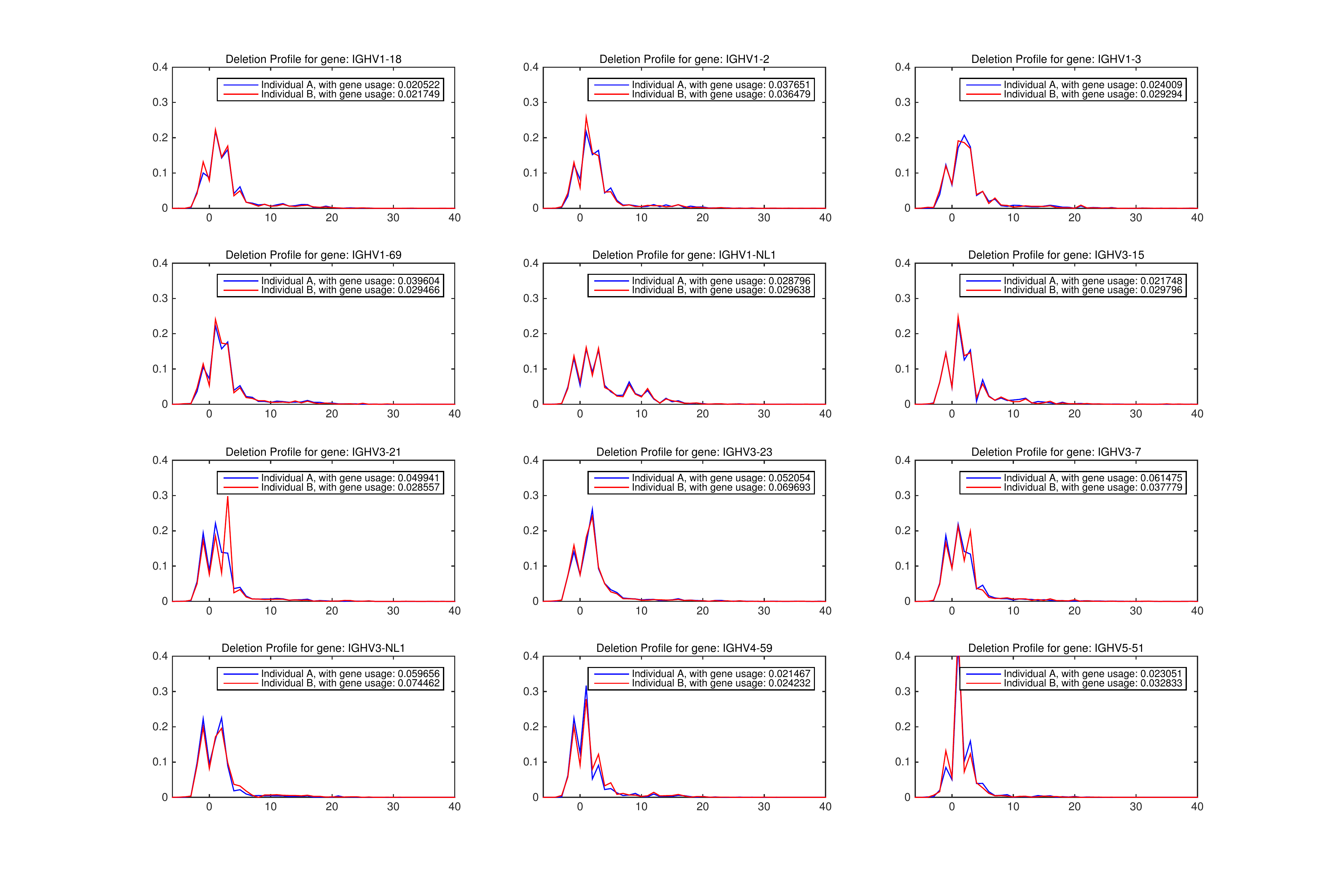}
\caption{Gene specific deletion profiles for a few frequently used V genes for individuals A and B (the profile for each of the two individuals is displayed separately on the same plot). The deletion apparatus is very sensitive to sequence context, and the deletion profile is quite variable from gene to gene. Remarkably, the deletion profile is nearly identical between the two subjects.}
\label{S4}
\end{center}
\end{figure*}

Although the deletion profiles are in fact gene dependent, and we infer a separate deletion profile for each gene (Fig.\,\ref{S4}),
for convenience we present here only the weighted mean over all genes of the V and J gene deletion distributions (Fig.\,\ref{RecomEvDist}c and d). These distributions seem little affected by selection, as evidenced by the similarity between the pre-selection and post-selection (naive) distributions.

\begin{figure}
\begin{center}
\includegraphics[width=\linewidth]{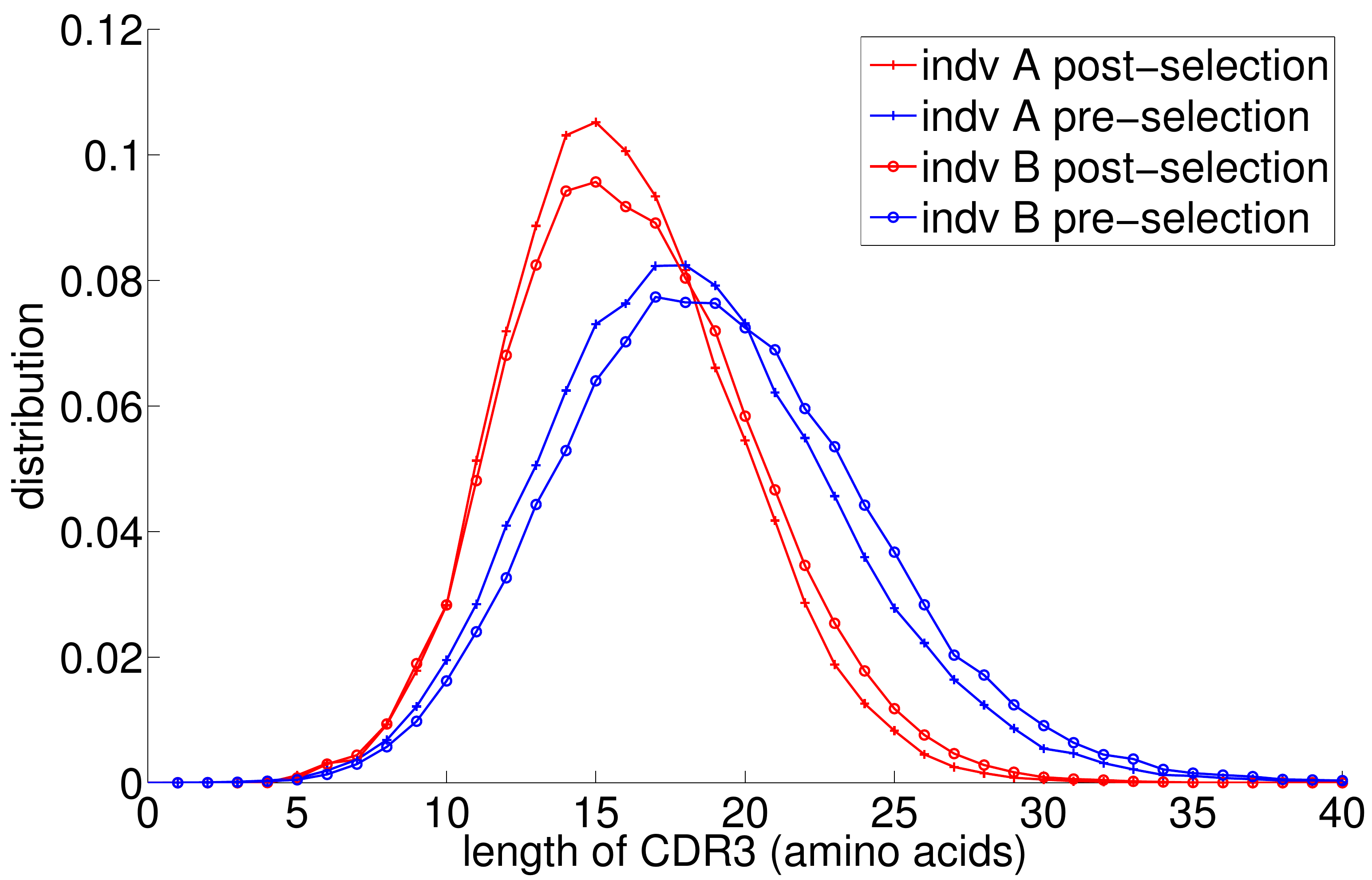}
\caption{Length distributions (in amino-acids) of the CDR3 region for different repertoires. The post-selection distributions are derived from the productive sequences in the naive repertoires. The pre-selection distribution is derived from a synthetic repertoire of productive sequences drawn from the generative model $P_{pre}$ that has been inferred from naive unproductive data sequences. Notable features include the progressive shortening and narrowing of the distribution as selective pressure is applied, and the close similarity, but not identity, between the two individuals.}
\label{LenDis}
\end{center}
\end{figure}

\subsection{Selection}


Armed with the raw recombination model $P_{\rm pre}$, we can estimate the effect of selection in the naive repertoire by comparing the statistics of the naive, productive sequences, with those predicted by the generation model.
The effect of selection is already evident from the CDR3 length distribution \cite{Harlan12}, as illustrated by Fig.\,\ref{LenDis}.
The pre-selection sequences are longer and have a wider length distribution than the selected ones. This effect of selection on length is in agreement with previous studies \cite{Harlan12,Shiokawa99}, and in close parallel with recent observations on T-cell receptors \cite{Elhanati08072014}.


\begin{figure}
\begin{center}
\includegraphics[width=\linewidth]{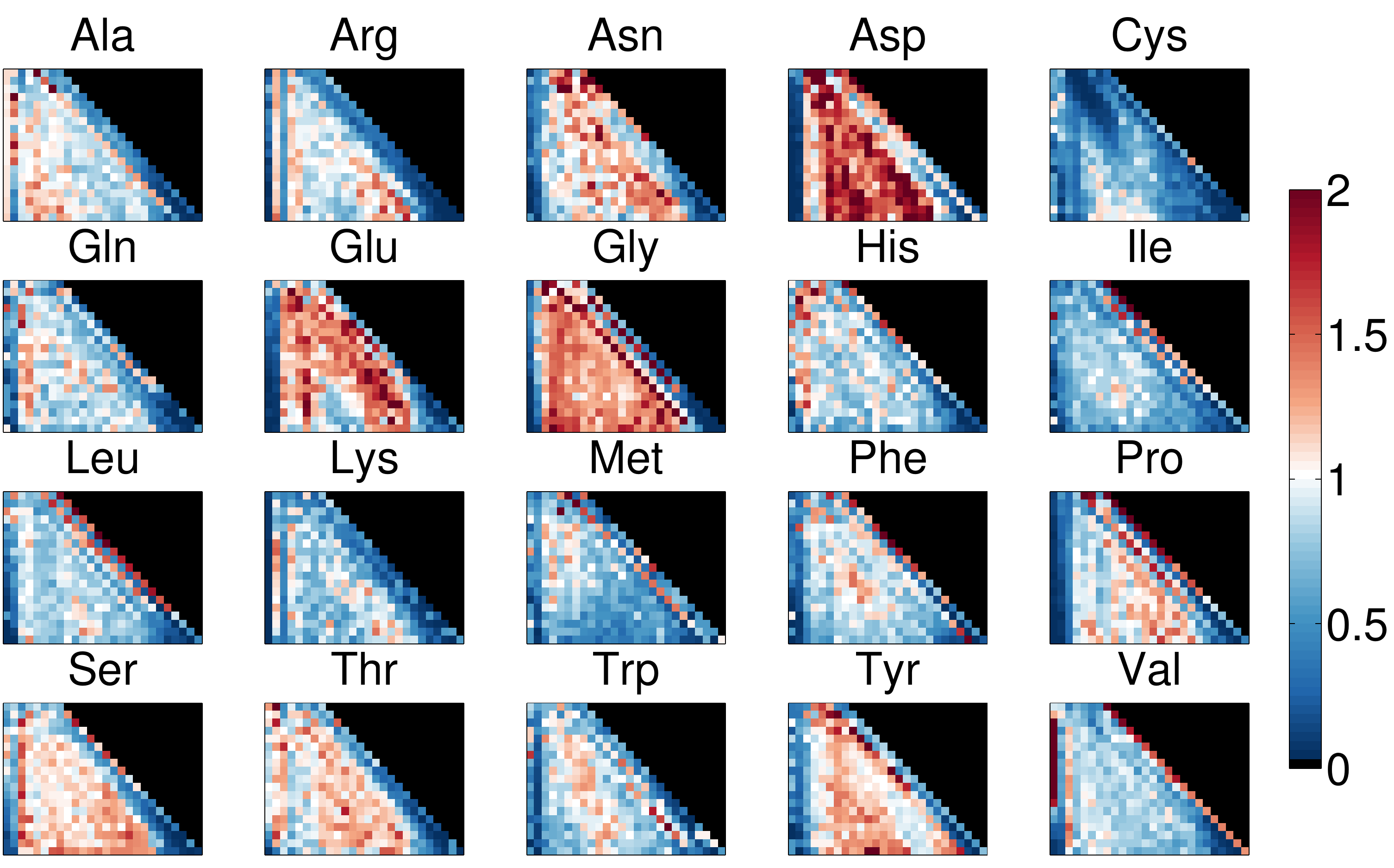}
\caption{Heat plot of the inferred amino-acid selection factors $q_{i;L}$ for each amino acid, ordered by length $L$ of the CDR3 region (ordinate) and position $i$ within that region (abscissa). The CDR3 region is bounded on the left by Cys residue and by a Trp residue on the right. There is a clear pattern of amino acid preference (or anti-preference) within a few positions of these boundaries, independent of overall CDR3 length $L$.}
\label{q1Heat}
\end{center}
\end{figure}

\begin{figure}[h!]
\begin{center}
\includegraphics[width=\linewidth]{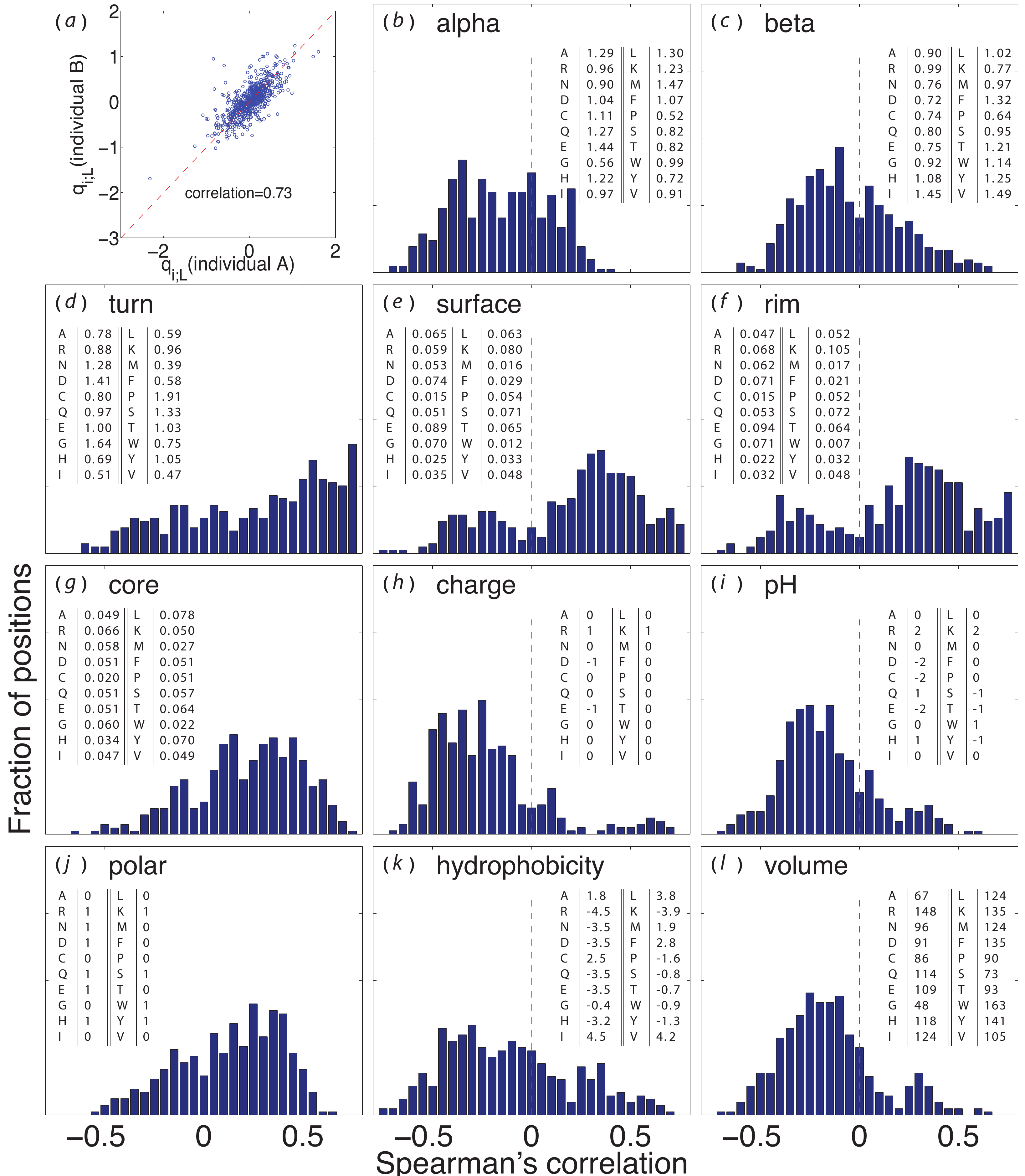}
\caption{({\bf a}) Scatter plot of the logarithms of the amino acid selection factors $q_{i;L}(a)$ between individuals A and B. The selection factors for the two individuals are strongly, if not perfectly, correlated. This justifies a joint analysis of the properties of the those factors, as done in the following panels ({\bf b-l}), showing correlation of the selection factors with several biochemical properties. Each panel shows the histogram, over all positions and lengths of both individuals, of Spearman's correlation coefficient between the selection factors for a given amino acid and the biochemical properties of that amino acid. The following biochemical properties are considered (from left to right, top to bottom): preference to appear in alpha helices ({\bf b}), beta sheets ({\bf c}), turns ({\bf d}) (source for ({\bf b-d}): Table 3.3 \cite{Stryer}). Residues that are exposed to solvent in protein-protein complexes (following definitions and data from \cite{MartinLavery}) are divided into three groups: surface (interface) residues that have unchanged accessibility area when the interaction partner is present ({\bf e}), rim (interface) residues that have changed accessibility area, but no atoms with zero accessibility in the complex ({\bf f}) and core (interface) residues that have changed accessibility area and at least one atom with zero accessibility in the complex ({\bf g}). Finally we plot the basic biochemical amino acid properties (source: {\tt http://en.wikipedia.org/wiki/Amino\_acid} and {\tt  http://en.wikipedia.org/wiki/Proteinogenic\_amino\_acid}): charge ({\bf h}), pH ({\bf i}), polarity ({\bf j}), hydrophobicity ({\bf k}) and volume ({\bf l}). For all properties the actual numerical values used to calculate the correlations are listed in the inset tables. 
}
\label{BioCorr}
\end{center}
\end{figure}

Selection acting on the rearranged sequences is quantified by position-dependent selection factors $q_{i;L}(a)$ (Fig.\,\ref{analysis_cartoon}b), which capture the positive or negative effect that each amino acid at each position has on the functionality of the entire sequence.
We show the amino acid selection factors averaged over both individuals in Fig.\,\ref{q1Heat}. 
These factors did not vary much between the two individuals (Fig.\,\ref{BioCorr}a), even though their initial repertoires were slightly different (including different length distributions, see Fig.\,\ref{LenDis}).
Residue selection patterns show a dependence on the position and length of the CDR3---some patterns are related to either the V or J side of the junction, while other effects are localized in the bulk. This is related to the conserved motifs just outside of the CDR3, which function as anchors for the variable area. Thus, the role of an amino acid is different whether it is close to the V gene conserved motif, the J conserved motif, or far from both. 

\begin{figure}
\begin{center}
\includegraphics[width=\linewidth]{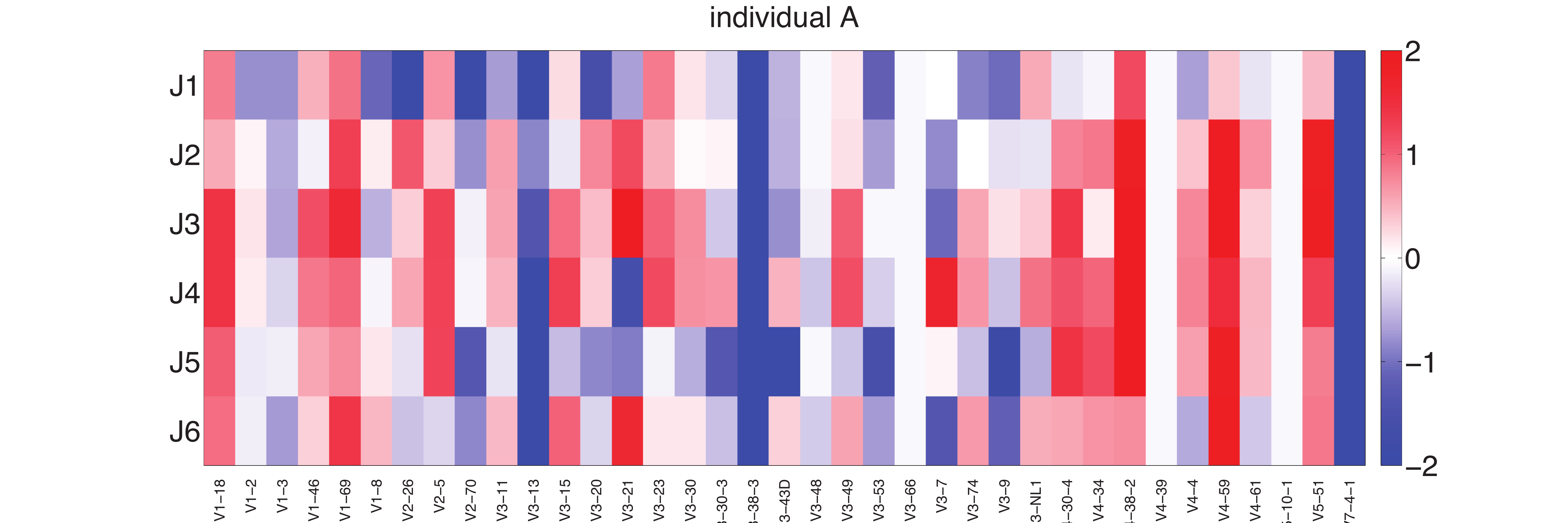}
\includegraphics[width=\linewidth]{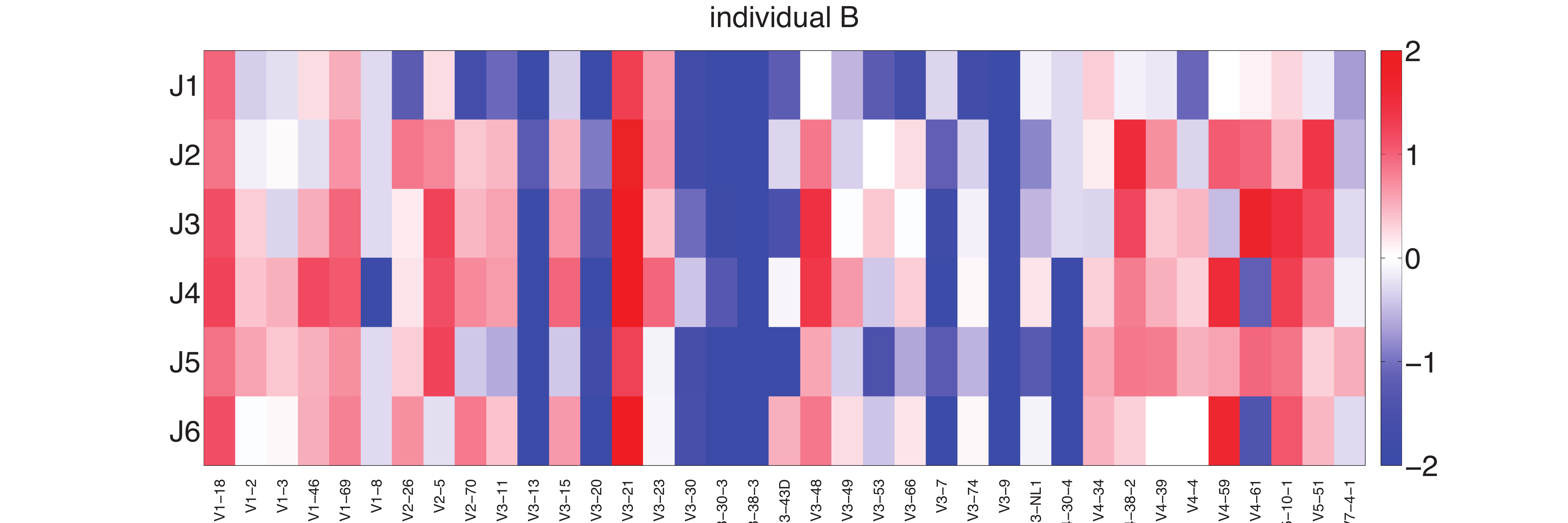}
\caption{Selection factors of pairs of V and J genes for both individuals (heat plot of the logarithm of the selection factors $q_{VJ}$). Again, only V genes with usage frequency greater than $.01$ for both individuals are shown. V selection pattern is different between individuals, but J pattern is more similar across genes.}
\label{S3}
\end{center}
\end{figure}


We also looked at correlation between our selection factors and various biochemical properties of the amino acids. We used quantified numeric properties for every amino acid, such as hydrophobicity or polarity, to look at the Spearman's correlation between these properties and selection factors of the 20 amino acids, for a specific position and CDR3 length. This can be done for every position and length, yielding many correlation coefficients. Figs.\,\ref{BioCorr}b-l show the distributions of those correlation coefficients. 
Selection is not determined by a single property of the amino acid.
However, some properties do correlate with selection, namely the tendency of an amino acid to participate in a turn of the protein (Fig.\,\ref{BioCorr}d) and its tendency to be found in the core of the interaction complex (Fig.\,\ref{BioCorr}g). A few other properties, namely amino acid volume (Fig.\,\ref{BioCorr}l), charge (Fig.\,\ref{BioCorr}h), pH (Fig.\,\ref{BioCorr}i) or hydrophobicity (Fig.\,\ref{BioCorr}k) all have a negative influence on the overall amino acid selection probability. This last negative correlation is consistent with the observation that hydrophobic D segments are selected against after rearrangement \cite{Harlan12}.
Most of these results are similar to what was observed in the case of T cells \cite{Elhanati08072014}.

\subsection{Correlation between generation and selection}
\begin{figure}
\begin{center}
\includegraphics[width=\linewidth]{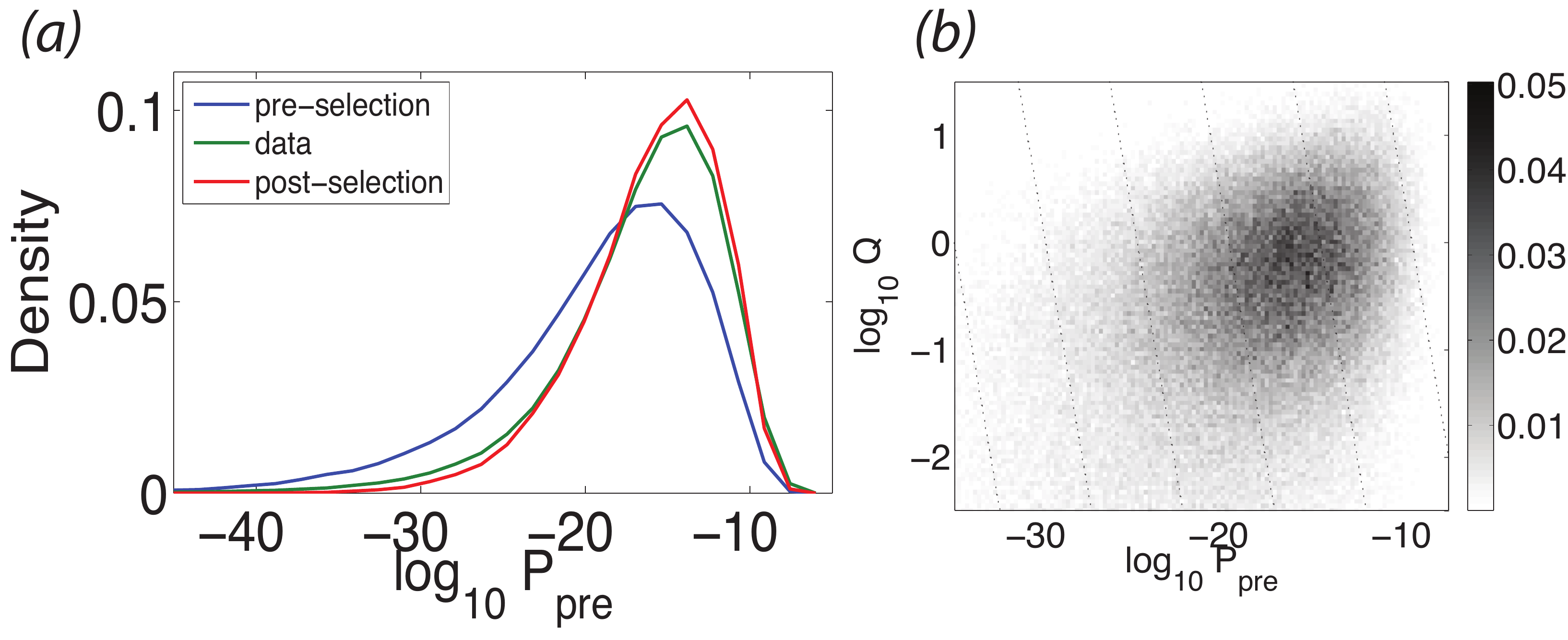}
\caption{Panel ({\bf a}) shows the distribution of generation probabilities (as inferred from the pre-selection model $P_{pre}$) for the preselection model itself (blue), the post-selection model $P_{post}$ (red), and the naive functional sequence repertoire itself (green). The key feature is that sequences in the selected repertoire have systematically higher generation probability. Panel ({\bf b}) makes the same point via a scatter plot of the primitive generation probability versus the selection factor $Q$ for a synthetic repertoire of sequences generated according to $P_{pre}$.}
\label{Pgen_Distn}
\end{center}
\end{figure}

What kind of sequences are likely to pass initial selection? Each sequence, no matter in what repertoire we observe it, can be assigned a probability of being generated in the initial VDJ recombination event. Fig.\,\ref{Pgen_Distn}a shows
the distribution of this quantity for sequences in the pre- and post-selection repertoires.
Remarkably, we note that most sequences have a very low generation probability (typically $<10^{-10}$).
The similarity of the naive productive (green) and post-selection model prediction (red) curve is a validation of the model, while the difference from the pre-selection (blue) curve highlights the effect of selection. Sequences that had higher probability to be generated are also more likely to be selected, resulting in a shift towards higher generation probabilities after selection.
This correlation between generation probability and selection is made more evident by
Fig.\,\ref{Pgen_Distn}b, which shows a 2D density plot of the generation probability and the overall selection factor $P_{\rm pre}$ versus $Q$, evaluated over a large set of generated sequences. There is a clear correlation between generation and selection---higher values of generation probability imply also higher values of selection and vice versa. To put it differently, the generation process anticipates the subsequent somatic selection process.

\subsection{Repertoire entropy}
\begin{figure}
\begin{center}
\includegraphics[width=\linewidth]{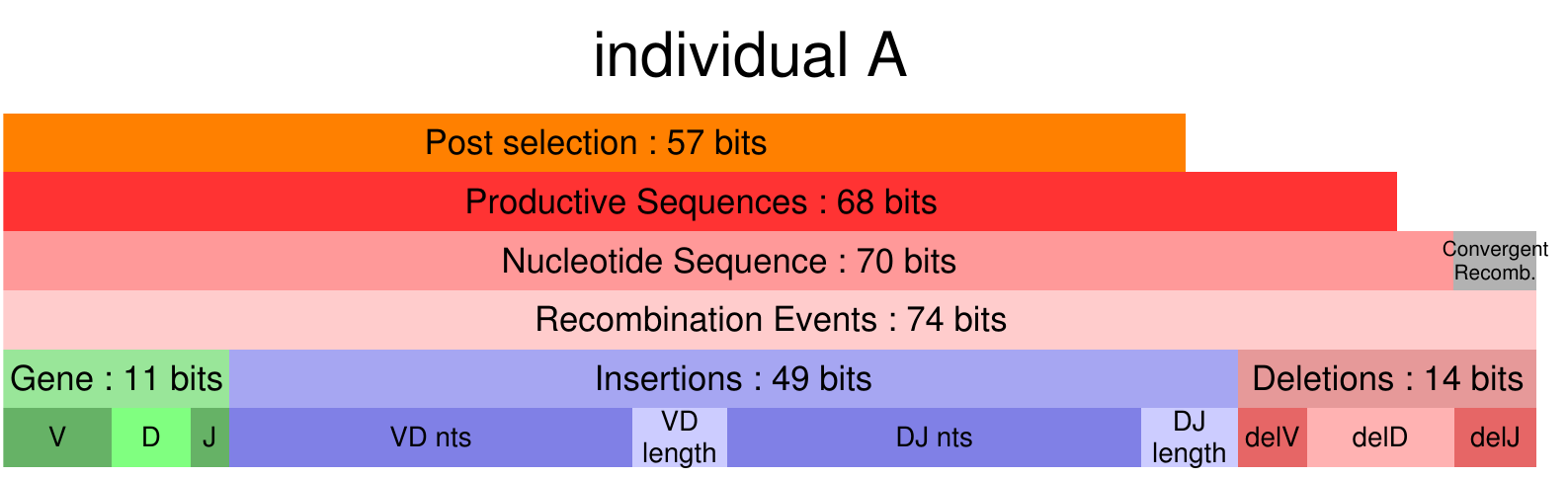}
\includegraphics[width=\linewidth]{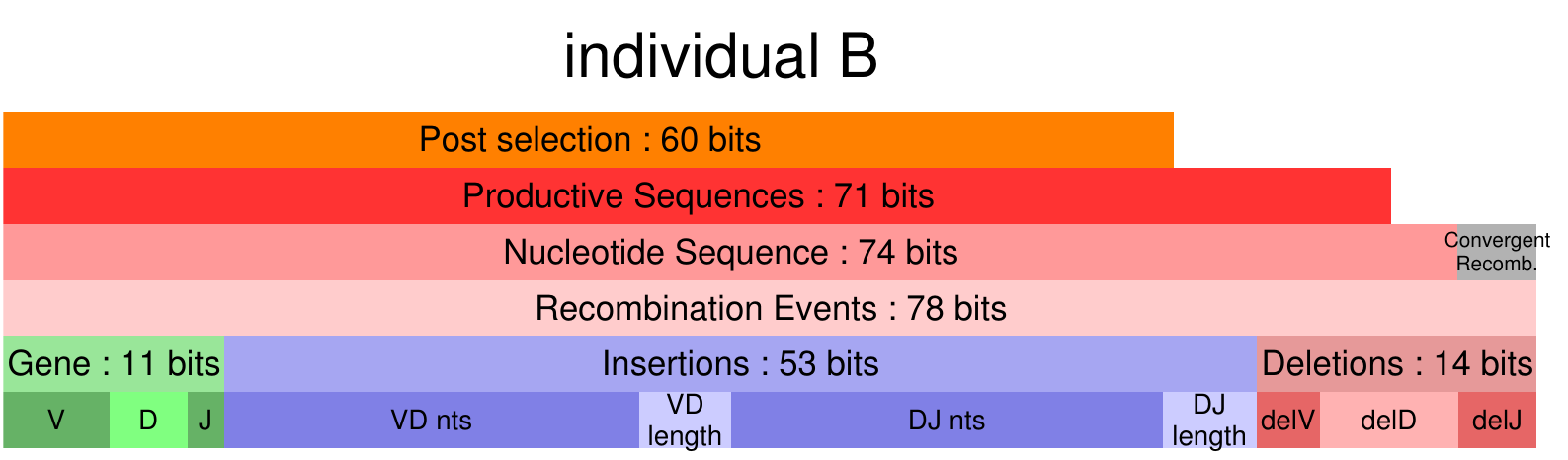}
\caption{Total sequence entropy partitioned into its various elementary contributions for the two individuals. The bottom three horizontal bars in each stack display the partitioning of the entropy of the probability distribution of recombination scenarios. Because multiple scenarios can generate the same sequence, the nucleotide sequence entropy of the sequences directly produced by recombination is smaller than the recombination scenario entropy. Out of those, productivity of the sequence further restricts the diversity by constraining frame and forbidding stop codons appearing, as depicted in the smaller bar above. Finally, as seen in the top most bar, the initial selection process itself significantly reduces the   diversity of those productive sequences. It is worth noticing that while the initial diversity of both individuals is different, consistent with their different CDR3 length distributions, the reduction effect of the selection is quite similar, keeping the same difference in entropy.}
\label{Entropy_partition}
\end{center}
\end{figure}

The diversity of the immune repertoire is one of its key characteristics.
The rearrangement and selection models enable us to precisely quantify this diversity and identify its sources.
Repertoire diversity is an inherently dynamic property. Upon random rearrangement, the initial diversity is established, but initial selection will modify it.
Those changes can be demonstrated by looking at the entropy of the different distributions, calculated from the models we infer. Entropy gives a measure of the number of different sequences we can expect to find at different stages of B cell development 
(Fig.\,\ref{Entropy_partition}). The generation entropy can be broken down into contributions from the different events that make up the recombination scenario. Most of the diversity comes from insertions. Note that the entropy of rearranged sequences is smaller than the entropy of recombination events. This is due to convergent recombination---the fact that a given sequence can be produced by different recombination scenarios. Individual B has larger generation entropy than individual A, primarily because it has more insertions. The entropy of productive sequences is further reduced (by 2 bits) by keeping only the in frame sequences.
Subsequent action of selection reduces the diversity of the repertoire by about 10 bits. This reduction is true for both individuals, regardless of their different generation entropy, and follows from the correlation between $P_{\rm pre}$ and $Q$ (Fig.\,\ref{Pgen_Distn}b), which concentrates the distribution towards the most likely rearrangements, thus reducing diversity in the selected repertoire. 

The numbers that can be associated with these entropies are extremely large: $2^{71}\sim 2\cdot 10^{20}$ for the productive sequences and $2^{60}=10^{18}$ for the naive repertoire -- much larger than the number of B cells in the body. These numbers are not estimates of the total number of {\em possible} sequences---this is set by the maximum number of insertions and is much larger---but rather the equivalent number of outcomes in a uniform probability distribution.



\subsection{Hypermutations}
Upon recognition of an antigen, B-cell receptors undergo an affinity maturation process, by which their binding strength to the antigen is increased through the combination of random somatic hypermutations and selection.
Thus, receptor sequences from antigen-experienced cells, such as memory cells, are expected to show the effect of somatic hypermutations, and we can use these sequences from our dataset to learn their statistics.
Hypermutations appear in our sequence reads as mismatches with the genomic sequence. However, because the survival of a sequence in the memory repertoire depends on its affinity for a particular antigen, the statistics of its hypermutations should reflect other factors than just the hypermutation process itself.
To overcome this issue, we make the assumption (as in \cite{Matsen14}) that when the hypermutation machinery is activated in a cell, it acts on both chromosomes indifferently. This assumption is backed by the comparable number of mismatches seen in the out-of-frame and in-frame memory data (the number of mismatches found in the naive out-of-frame data is smaller by two orders of magnitude).  If this assumption is true, out-of-frame sequences will also display the effects of somatic hypermutations, with statistical properties unaffected by any further selection effect. 
For this reason, we restrict to out-of-frame memory sequences to infer the statistical properties of the somatic hypermutation process.

\begin{figure}[h!]
\begin{center}
\includegraphics[width=\linewidth]{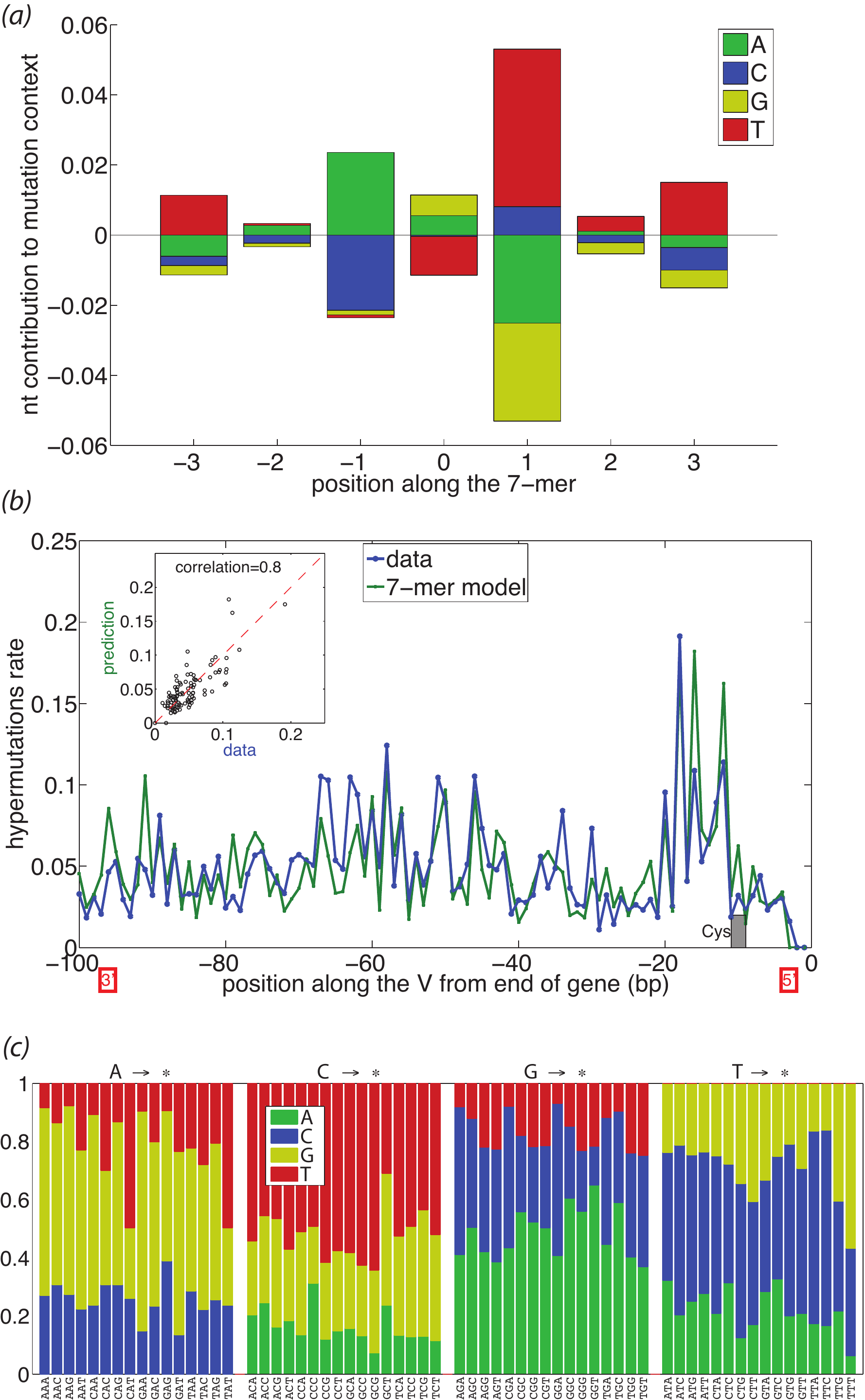}
\caption{Sequence dependence of somatic hypermutations. ({\bf a}) The model mutation probability depends on the central base (position 0) and on the sequence context, three base pairs on each side. The log relative probability of a mutation is the sum of contributions (positive and negative) read off from the sequence motif according to the sequence at each of the seven positions. ({\bf b}) Comparison of the predictions of this model with the observed hypermutation rate at different positions within the V gene. Mutation \lq hotspots\rq\ are well predicted, and the scatter plot (inset) between data and prediction shows strong correlation. The location of the Cys anchor of the CDR3 is indicated, and we note that the hypermutation rate (in data and model) is low within this special codon. ({\bf c}) Substitution probabilities to the different bases as stacked columns vs. the local trimer context, grouped by the central base. Substitution is not uniform, depending primarily on the base being mutated, but varying with the context.}
\label{hotspot_plots}
\end{center}
\end{figure}

First, we used the out-of-frame memory sequence reads to infer a model of rearrangement with random hypermutations in the germ line part of the sequence (Fig.\,\ref{analysis_cartoon}b).
Such a model allows us to assign a set of likelihood-weighted recombination and hypermutation scenarios to any specific sequence.
We thus identified and probabilistically weighted somatic hypermutation events and recorded them, together with their sequence context within a 7-nucleotide window centered on the mutation. We restricted our analysis to V gene-derived segments: D and J gene derived sequences are much shorter and may not have provided enough statistics for our analysis.
We then use statistics of the sequence context of hypermutations to construct a simple additive scoring model for the probability of mutation at any position within a V gene context. Specifically, the probability of observing a somatic hypermutation was assumed to be of the form $p_{\rm SHM}(\sigma) \propto p_{\rm bg}(\sigma)\exp[\sum_{i=-3,3}e_i(\sigma_i)]$ where $\sigma = (\sigma_{-3},\sigma_{-2},\ldots,\sigma_{+3})$ is the nucleotide sequence in a 7-nucleotide window around the mutated nucleotide $\sigma_0$; $p_{\rm bg}(\sigma)$ is the background probability of the heptamer $\sigma$ occurring in the genomic V gene sequences; and the $e_i(\sigma_i)$ are contributions of each position to the motif, which play the same role as binding energies \cite{Berg1987723}. The $e_i$'s are adjusted so as to maximize the likelihood of the data under the model.
The $e_i$ factors so inferred are displayed in Fig.\,\ref{hotspot_plots}a. Since they are defined up to a constant, we impose $\sum_{\sigma} e_i(\sigma)=0$ at each position $i$ relative to the mutation site.
A positive value of $e_i(\sigma)$ means that nucleotide $\sigma$ is enriched at position $i$ relative to a mutation site.

By running $p_{\rm SHM}(\sigma)$ over the genomic V gene sequences, we get a predicted probability of observing a somatic hypermutation at different positions within the V gene.
The prediction of this model averaged over all gene profiles (aligned with respect to the conserved Cys),
and its comparison with observed hypermutations is displayed in Fig.\,\ref{hotspot_plots}b
(the profiles for individual genes were found to be similar to the average profile).
The scatter plot of model predictions vs data (correlation = 0.8) shows that the model gives a good account of the data, and the plot of their position dependence shows that the model accounts well for the presence of hypermutation hotspots. Note that the hypermutation rate displayed in Fig.\,\ref{hotspot_plots}b is very large: 5 to 10\% per position. 
For comparison, the rate of mismatches in the naive out-of-frame sequences (attributable to sequencing errors or a leak of memory cells into the naive set) was $\sim 0.1\%$.
Finally, we also looked at the dependence of the substituted nucleotides on the immediate context of the mutation. Fig.\,\ref{hotspot_plots}c represents the probabilities of substituted nucleotides as a function of the trimer context. 
We note a clear dependence on the identity of the mutated base, with additional context-dependent variability from
the local trimer sequence.

\section{Discussion}
Our statistical algorithms allowed us to characterize in detail the generation, initial selection and hypermutation processes that lead to the observed B-cell receptor repertoires. Two key ingredients underlie our approach. First, we exploited out-of-frame sequences, assumed to be free of selective pressure, to reconstruct the statistics of the DNA editing processes: VDJ recombination and hypermutations. Out-of-frame sequences have similarly been used as a baseline to study the properties of B-cell receptor naive repertoires \cite{Harlan12}, or to estimate selective pressures related to affinity maturation in memory B-cells \cite{Matsen14}.
Second, our approach overcomes the degeneracy of the recombination problem (whereby the same sequence may be generated by many different recombination events) by using a fully probabilistic approach.

The generation of B- and T-cell receptors results from similar processes, involving common enzymes and pathways. Thus, perhaps not surprisingly, many of the results we obtain here are similar to what was reported for T-cell receptors using similar methods \cite{Elhanati08072014}: statistical independence between the insertion and deletion processes, as well as between gene choice and insertions; the identity of inserted nucleotides following a Markovian probability law. 
Another similarity with T-cell receptors is that the generation process anticipates the action of selection: sequences that are more likely to be produced are more likely to be retained by selection. This suggests evolutionary adaptation of the generation machinery.

We also noted some differences with T-cell receptors.
Because B-cell receptors have more insertions than TCRs, we find that B-cell repertoires are much more diverse than T-cell repertoires, as measured by entropy, even before hypermutations occur. In addition, the selection factors acting on the CDR3 amino-acid sequence of B-cell receptors are quite different from those reported for T-cell receptors,
consistent with the fact that their cognate epitopes are very different in nature.


Although there is a difference in diversity between our two individuals, this difference is restricted to the generation process, and is caused by a slight excess of insertions in individual B.
Selection factors on the other hand seem to be well conserved across individuals, pointing to general biophysical and biochemical properties that are subject to selection.
The selection factors correlate with some biochemical properties, including a negative correlation with hydrophobicity (in agreement with a previous report that hydrophobic D segments are selected against \cite{Harlan12} after rearrangement).


This study also provides a couple of methodological advances.
Thanks to our stochastic framework, we were able to identify heterozygous genes as well as their alleles, even when these were not present in existing databases. We could then reconstruct the partition of these alleles into two chromosomes. We found no significant bias in chromosome usage. 

Perhaps the most important difference between B- and T-cell receptors is the existence of hypermutations accumulated during affinity maturation. Hypermutations are expected to add an enormous amount of sequence diversity. With a hypermutation rate estimated to be of the order of 5-10\%, hypermutations are expected to contribute an additional $\approx 0.4$ bit per nucleotide---a huge number if we consider that hypermutations can happen over a region a few hundred nucleotides in length.
A common difficulty in studying hypermutations is to disentangle the biases that are inherent to the hypermutation process from the biases that result from affinity-driven selection \cite{DunnWalters98}. While previous efforts have been devoted to the inference of selective pressures in the B-cell memory repertoire \cite{Matsen14}, here we focused our attention on the raw substitution process.
We used out-of-frame memory sequences to learn about the statistics of substitutions unperturbed by the effect of selection, as was suggested in \cite{Dunn-Walters01031998}. For this purpose, we adapted our probabilistic framework to infer the hypermutations profile across all genes.
The hypermutation rate was found to be very variable along the V-gene sequence but similar across genes.
We showed that a simple additive model 
could correctly predict the hypermutation rate from the immediate sequence context (heptamer) around the mutation site. These results confirm earlier reports that hypermutations are context dependent both in localisation and substitutions \cite{Dunn-Walters01031998,Cowell15022000,Oprea15012001,Spencer15102005}, as well as recent observations from high-throughput sequencing data restricted to synonymous mutations \cite{Yaari13}. Our additive model is consistent with an independent site binding energy model for the hypermutation-inducing enzyme AID, where each flanking nucleotide contributes independently to its binding energy, as in the case of transcription factor binding sites \cite{Berg1987723}.

The hypermutation rates and substitution matrices inferred by our model could serve as a baseline or a control for future efforts to infer selective pressures from functional sequences, and could be useful in the inference of mutational phylogenic trees in the study of affinity maturation.



{\bf Acknowledgements.} We thank Harlan Robins for the BCR data without which this work would not be possible. We also thank Mr. Shai Chester for assistance in developing our allele-finding algorithms. The work of YE, QM, TM and AW was supported in part by grant ERCStG n. 306312. The work of ZS and CC was supported in part by NSF grants PHY-0957573 and PHY-1305525.


\bibliographystyle{pnas}
\bibliography{Bcell_refs}

\end{document}